\documentclass[final,5p,times,twocolumn]{elsarticle}

\usepackage{lineno,hyperref}
\usepackage{wrapfig}

\modulolinenumbers[5]

\journal{Journal of \LaTeX\ Templates}

\begin{document}

\begin{frontmatter}

\title{
The TOP counter of Belle II: status and first results
}

\author{Umberto Tamponi\fnref{myfootnote}}
\address{INFN - Sezione di Torino, 10124 Torino}
\fntext[myfootnote]{tamponi@to.infn.it}

\begin{abstract}
High-efficiency and high-purity particle identification are fundamental requirements for the success of the Belle II experiment, whose main goal is to explore the new-physics scenarios in the CP-violating decays of the B mesons.
To achieve the required particle identification performances, the Time-of-propagation counter has been installed in the central barrel region. This unique device consists in 16 bars of fused silica that act simultaneously as radiator and as light guide for the Cerenkov light. Unlike in the DIRC detectors, the PID information is mostly extracted measuring the time of propagation of the Cherenkov light in the radiator rather than its purely geometrical patterns.
We will present here a general overview of the status of the TOP counter, including the estimation of the time resolution, the calibration strategies and performances, and the first result obtained in the commissioning phase, both using cosmic rays and $e^+e^-$ collision events collected during the {\it phase II}
pilot run of the Belle II experiment.
These are the first measurements of the particle identification performances of a time-of-propagation detector in a full HEP experimental setup.
\end{abstract}


\end{frontmatter}


\section{Overview}
The Belle II experiment  \cite{Abe:2010gxa} at the SuperKEKB collider aims to collect $50 $ ab$^{-1}$ of $e^+e^-$ collisions at the $\Upsilon(4S)$ and the nearby bottomonium resonances $\Upsilon(3S, 5S, 6S)$ to perform precision measurements of the rare $B$ meson decay, search for signatures of new physics in the dark sector, and study the spectroscopy of the exotic hadrons \cite{Kou:2018nap}.

The first part of the data taking, the pilot run called {\it phase II}, started in April 2018 and lasted until July, collecting a total luminosity of about $0.5 $ fb$^{-1}$. All the sub-detectors were installed during the data taking except for the inner silicon tracker, that was almost completely replaced by beam-background monitoring sensors. Only one eight of the inner tracker was installed for commissioning purposes. 
The phase II dataset has been used to commission the experiment, perform the early calibration and determine the initial performances of each sub-detector. In the following, we will discuss in detail the results of the commissioning of the Time-Of-Propagation (TOP) counter.

\section{The TOP counter}

The TOP counter of the Belle II experiment is the only existing, operational time-of-propagation Cherenkov counter \cite{Akatsu:1999hi,Ohshima:2000ep,Inami:2008zz,Inami:2014nra}, and {\it phase II} represented the first attempt to perform particle identification with such device in a collider experiment. 
\begin{figure}[ht!]
  \begin{center}
    \includegraphics[width=0.6\textwidth]{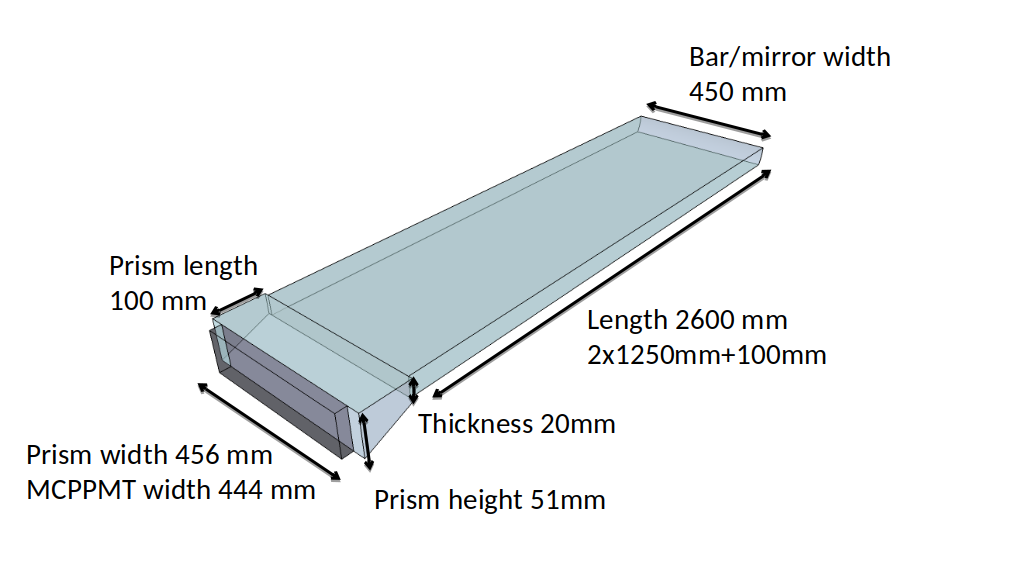}
    \caption{Sketch of one of the 16 modules of the TOP detector. The junctions between the two bar sections and between the bar and the mirror section are not shown.}
    \label{fig:topdesign}
  \end{center}
\end{figure}
\begin{figure*}[ht!]
  \begin{center}
    \includegraphics[width=1.\textwidth]{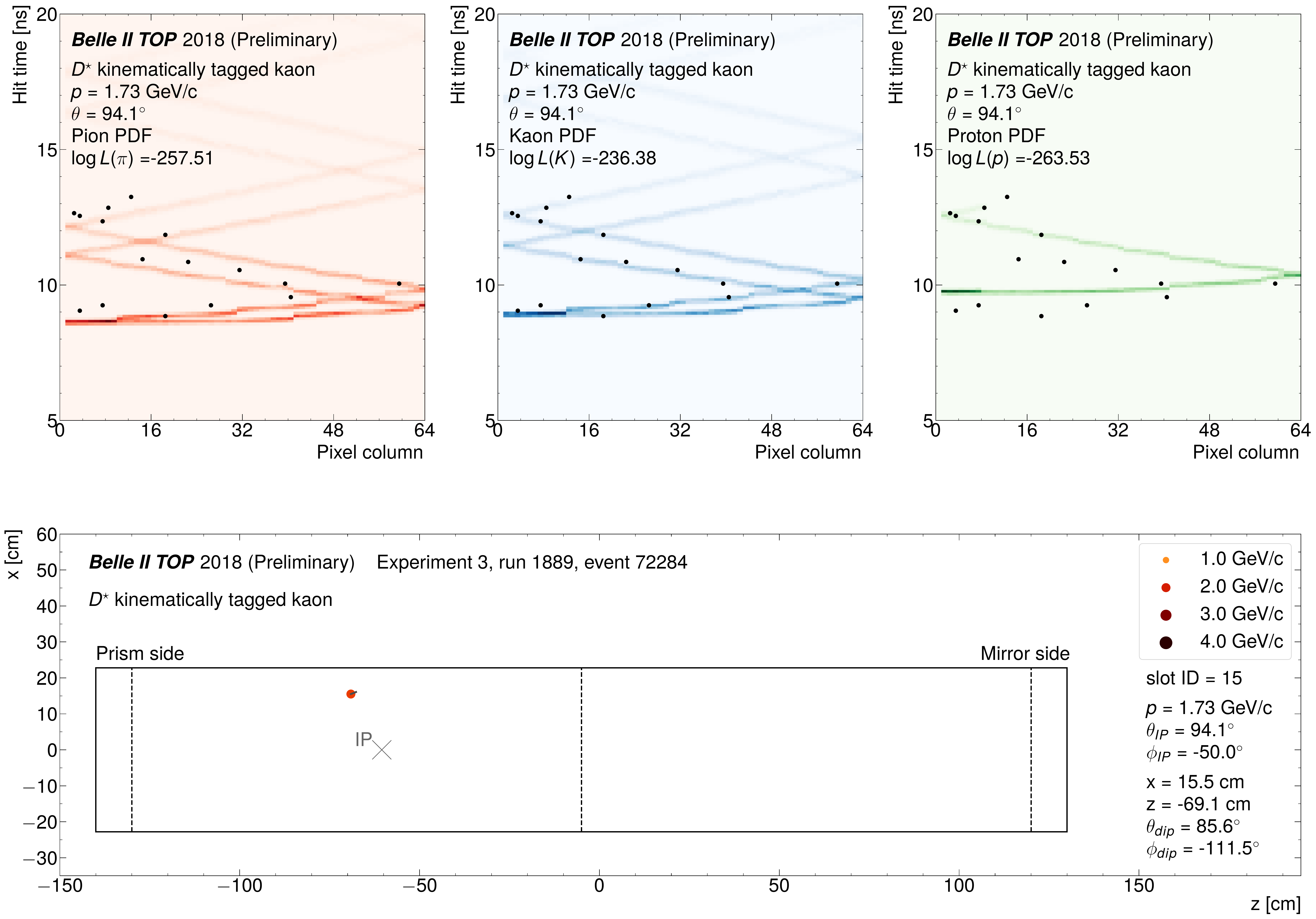}
    \caption{{\it upper panels}: space-time distribution of the hits associated to a kaon candidate track selected in the phase II data. The x-axis represents the position of the pixel along the bar transverse dimension, while the y-axis represents the detection time, referred to the most probable bunch crossing. The black points represent the observed hits, while the smooth distribution the expected PDF for a pion ({\it upper left}),  a kaon ({\it upper center}) or a proton ({\it upper right}) of same momentum.
        {\it lower panel} : reconstructed impact point and direction of the kaon candidate at its entrance in the TOP active volume.}
    \label{fig:event_9}
  \end{center}
\end{figure*}
It is composed by sixteen identical modules as the one sketched in Figure \ref{fig:topdesign}, arranged around the interaction point in a barrel-like geometry. Each module is composed of four parts glued together: two fused silica bars of dimensions ($125 \times 45 \times 2$) cm acting as Cherenkov radiator, a mirror located at the forward end of the bars, and a 10 cm long prism that couples the bar with an array of micro-channel-plate photomultiplier tubes (MCP-PMT) \cite{INAMI2012683,Matsuoka:2015knj}. Thanks to the high average refractive index ($n = 1.44$ at $405$ nm) of the fused silica at least part of the Cherenkov radiation emitted by the particles crossing the radiator remains trapped by total reflection, propagating to the MCP-PMT array. Having a pixel size of approximately $5.5 \times 5.5$ mm and a transit time spread less than $50$ ps, the MCP-PMTs provide a coarse measurement of the photon positions and a very precise measurement of their detection time. The photo-electron detection time, measured respect to the initial $e^+e^-$ collision, can be decomposed into two contribution: the time of flight of the charged particle from the interaction point to the TOP, and the time of propagation of the Cherenkov light inside the quartz. Once the direction of the incoming particle is known, the latter is function of the Cherenkov angle. The TOP therefore provides a combined measurement of both time of flight and Cherenkov angle. 

The particle identification information is extracted comparing the distribution of the time of arrival of the photons in each of the 512 channel with the expected  PDFs for six particle hypotheses ($e$, $\mu$, $\pi$,$K$, $p$, $d$) \cite{Staric:2011zza}. The six corresponding likelihood values are then stored, and their ratios are used to assign identification probabilities.

\section{First results using pure samples of kaons}
The TOP particle identification capabilities have been tested selecting pure samples of pions, kaons and protons tagged reconstructing the decay chains $D^{\star +} \to D^0 \pi^+ \to K^- \pi^+ \pi^+$, $K_s \to \pi^+ \pi^-$, and $\Lambda \to p \pi^-$ \footnote{Charge conjugation is understood for all the processes discussed in this paper.}. We will focus here on the kaon/pion separation power and on the pion fake rates, determined using the $D^{\star +}$ and the $K_s$ decays in the first $90$ pb$^{-1}$ of data. All the results presented here have been obtained with preliminary, severely limited time calibrations and without any geometrical alignment. This prevents us from presenting here precise numerical results.

The $D^{\star+}$ reconstruction begins with selecting the $D^0 \to K^- \pi^+$ candidates. The $D^0$ is reconstructed from track pairs of opposite charge pointing to the primary interaction point. One track is assigned with the $K$ mass hypothesis and the other with the $\pi$ one, without using any particle identification information. The kaon candidate is required to be within the TOP acceptance. 
After applying a kinematic fit to constrain the track to a common vertex, we discard most of the combinatorial background requiring the  $D^0$ candidate to have mass within $1.85$ GeV/c$^2$ and  $1.88$ GeV/c$^2$, corresponding to a  $2.5 \sigma$ window around the $D^0$ peak. 
The surviving $D^0$ candidates are then combined with an additional track with charge opposite of the kaon one to reconstruct the $D^{\star +}$ candidates. Again, we apply a vertex constrained kinematic fit and we require the mass difference between the $D^{\star+}$ and the $D^0$ candidate to be within $143.6$ MeV/c$^2$ and $147.6$ MeV/c$^2$.
Finally we further suppress the background requiring the $D^{\star+}$ candidates to have momentum in the center-of-mass frame greater than $2.5$ GeV/c$^2$. 
The result of this procedure is a small, but pure sample of $K$ with less than $5\%$ of contamination from other particles, mostly pions.

Using this pure kaon sample one can clearly visualize how the Cherenkov rings are reconstructed in a coordinate-time space by the TOP counter. Figure \ref{fig:event_9} shows the MCP-PMT hit timing distribution associated to a kinematically tagged kaon, compared  with the PDFs expected for a pion, kaon or proton of the same momentum and angle.

\begin{figure}[ht!]
 \begin{center}
   \includegraphics[width=.4\textwidth]{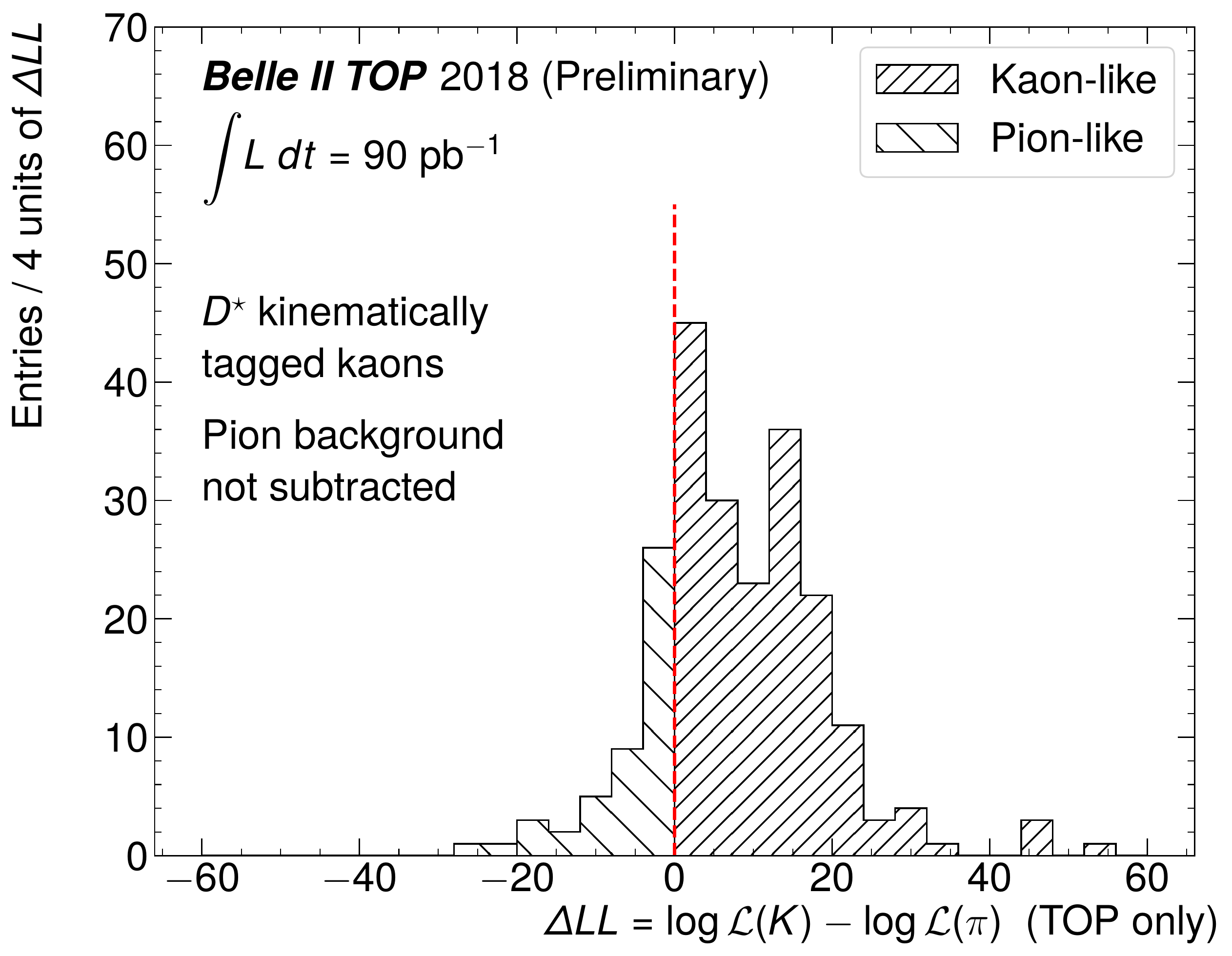}
    \caption{Difference between $\log{\cal L}_K$ and $\log{\cal L}_{\pi}$ for the kaons tagged by the $D^{\star+} \to D^0[\to K^- \pi^+] \pi^+$ decay. Only the TOP detector is used to calculate the likelihood values.}
    \label{fig:DeltaLL}
  \end{center}
\end{figure}

For each kinematically tagged kaon, we calculate the likelihood values ${\cal L}_{\pi}$ and ${\cal L}_{K}$ for the pion and kaon hypothesis by comparing the observed time and spacial distribution of the detected photons with the expected ones. Figure \ref{fig:DeltaLL} shows the log-likelihood difference  $\Delta LL = \log{\cal L}_K - \log{\cal L}_{\pi}$. The distribution is shifted towards positive values, indicating that the TOP is more likely to identify kaons as kaons rather than pions, as expected.

\section{First results using pure samples of pions}
To measure the probability ${\cal P}(\pi \to X)$ for a pion to be misidentified as another particle $X$, we reconstruct the $K_s \to \pi^+\pi^-$ decay, applying the same loose criteria used to select the tracks for the $D^{\star+}$ reconstruction. In addition, one of the two pions is required to be within the TOP acceptance ({\it probe}), while no selection is applied to the other track. We then study the yield of $K_s$ as a function of the TOP response for the {\it probe} pion.  Requiring $\log{\cal L}_X > \log{\cal L}_{\pi}$ we estimate  ${\cal P}(\pi \to X)$, while to measure ${\cal P}(\pi \to \pi)$ we require  $\log{\cal L}_{\pi} > \log{\cal L}_{K}$.
\begin{figure}[htbp!]
  \begin{center}
   \begin{tabular}{c}
    \resizebox{.73\columnwidth}{!}{\includegraphics{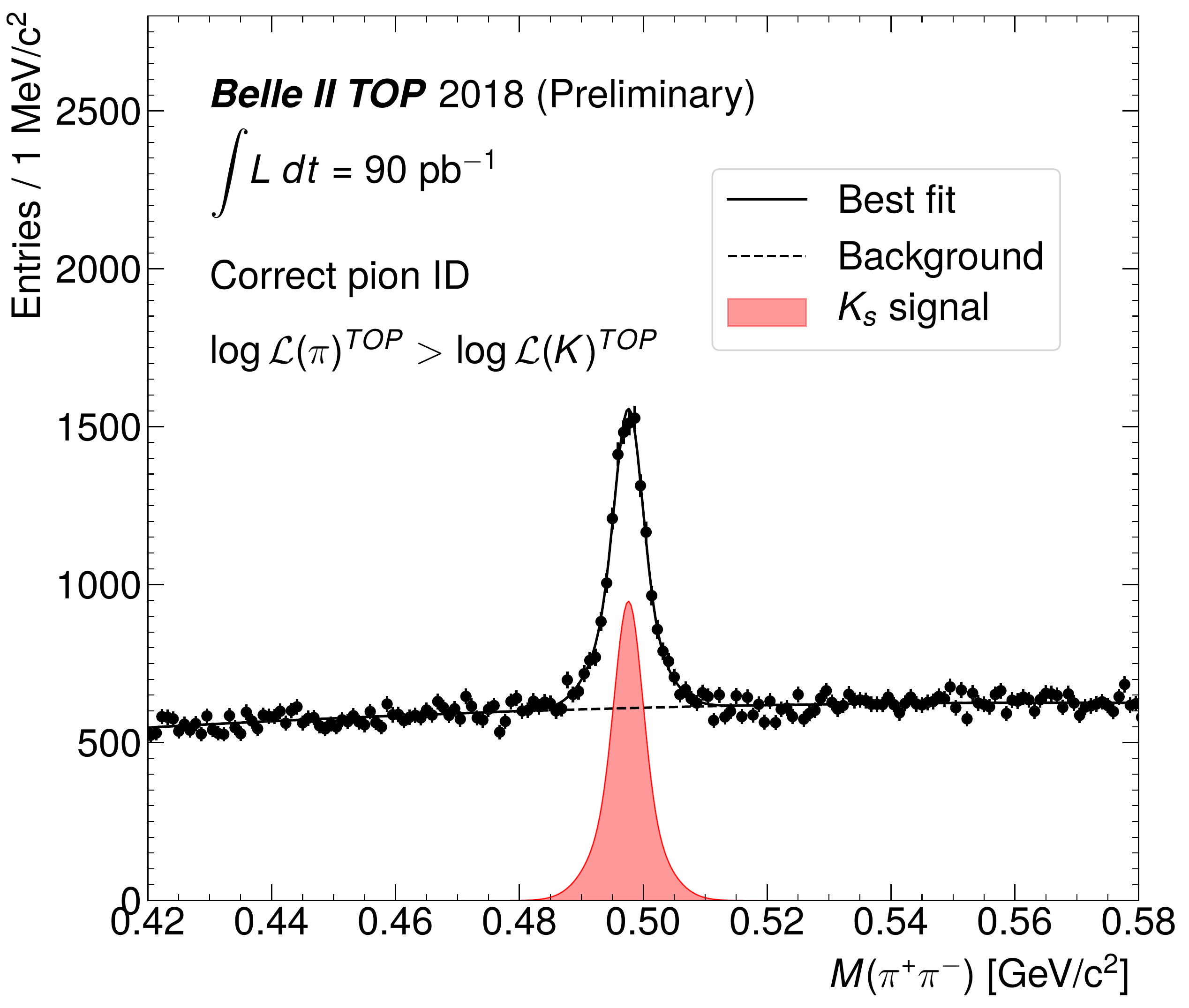}} \\
    \resizebox{.73\columnwidth}{!}{\includegraphics{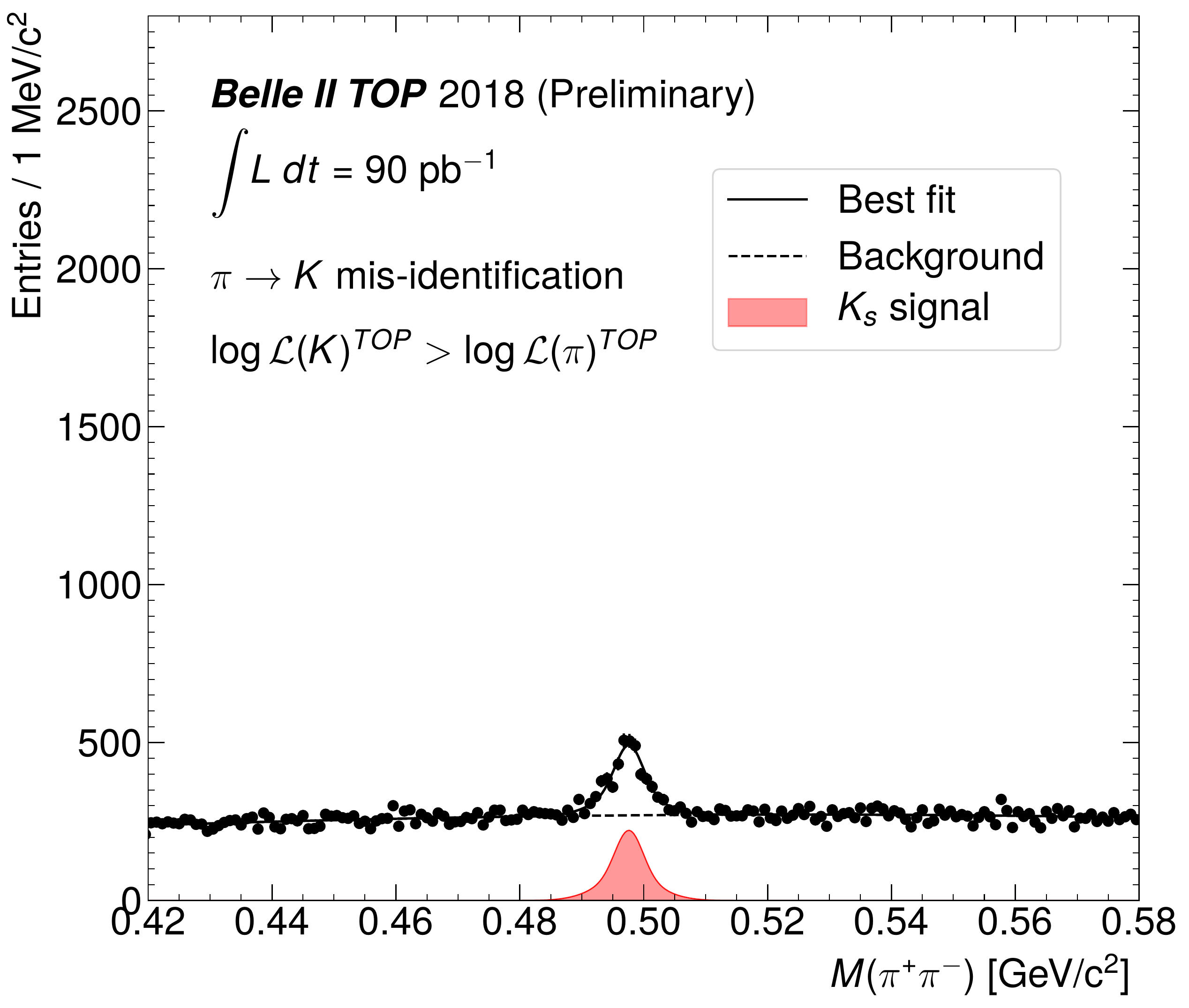}}  \\
    \resizebox{.73\columnwidth}{!}{\includegraphics{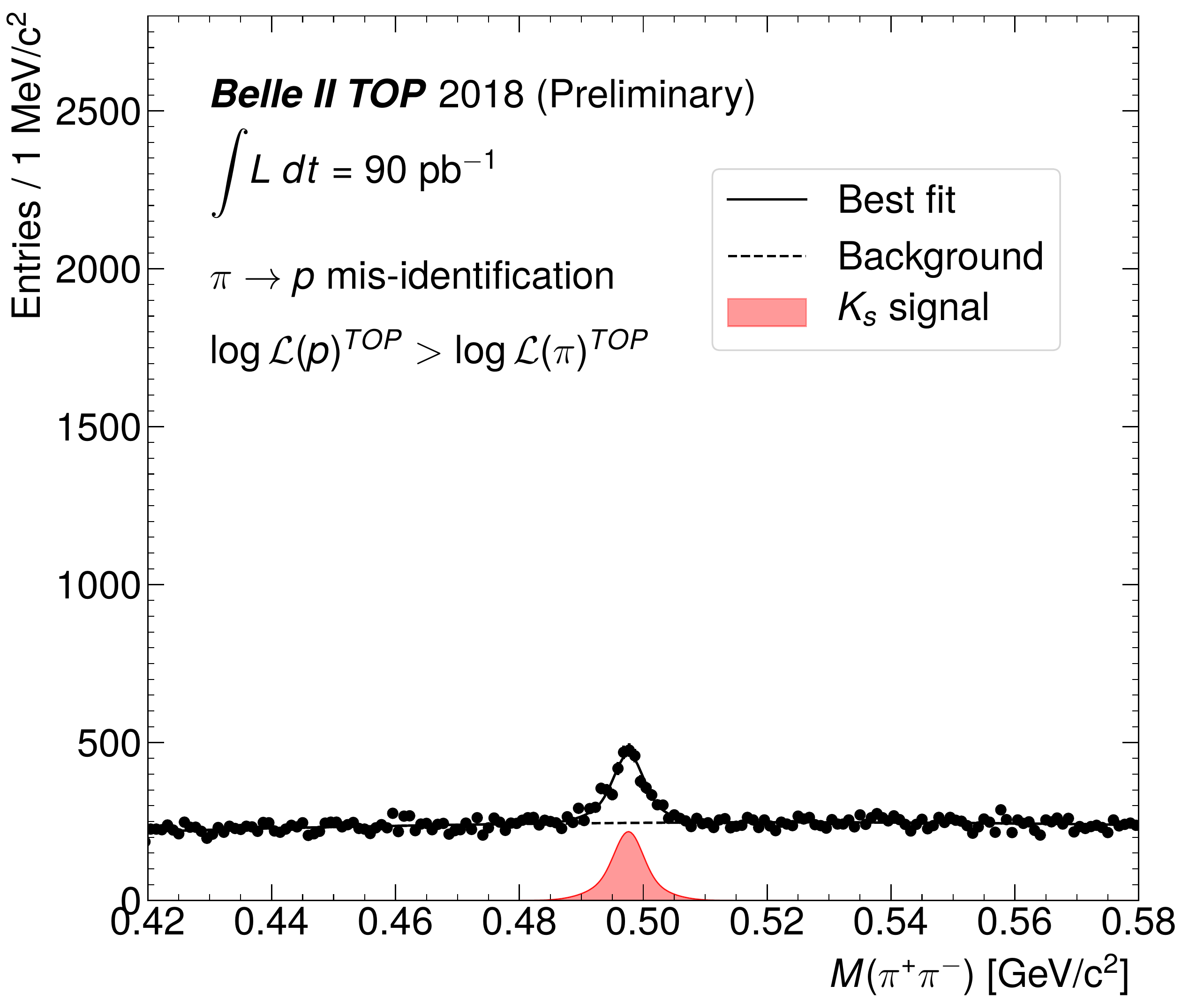}} \\
    \resizebox{.73\columnwidth}{!}{\includegraphics{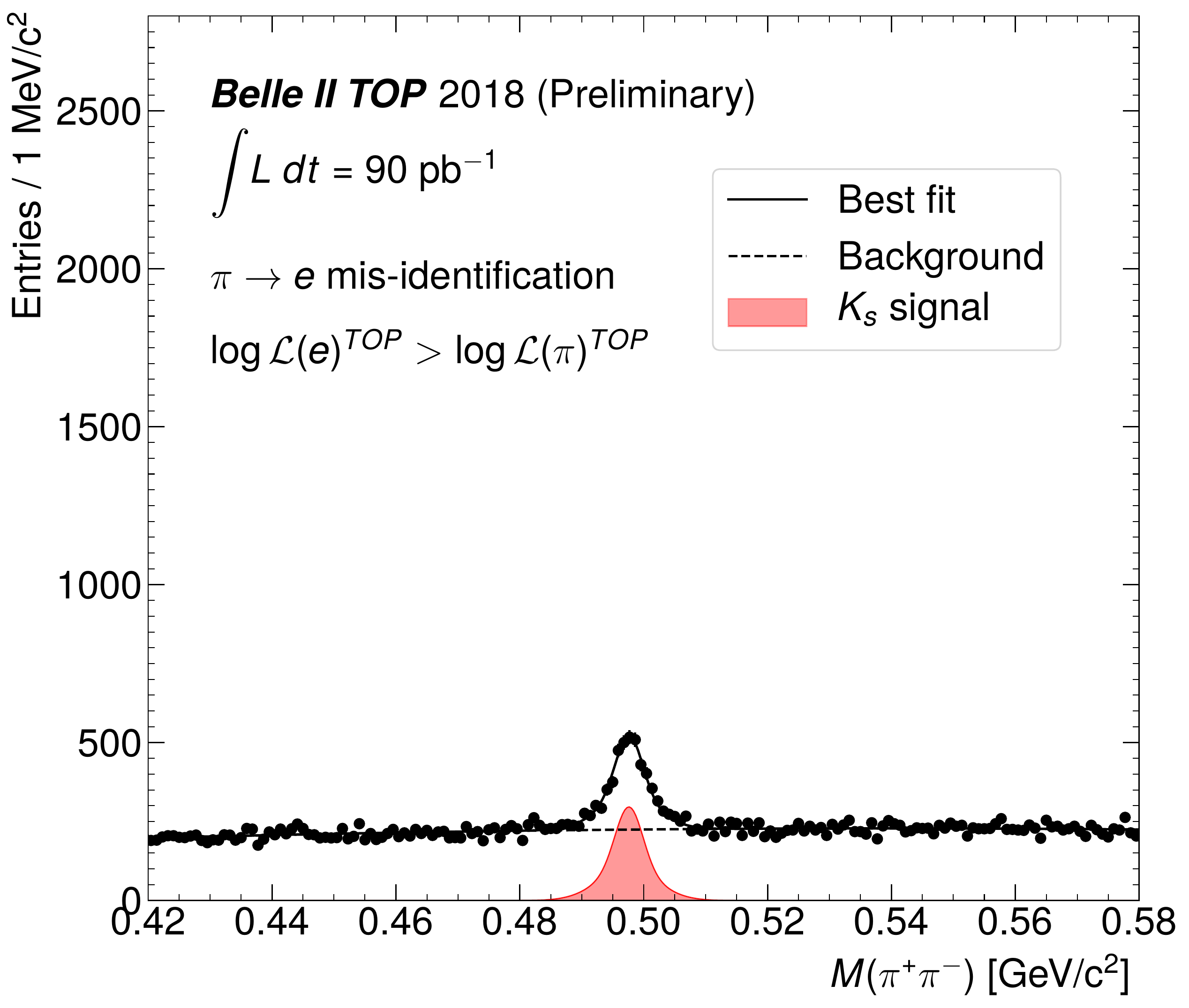}}  \\
    \end{tabular}
    \caption{ Mass distribution of  the $K_s$ candidates with four particle identification requirements. From top to bottom: $\log{\cal L}_{\pi} > \log{\cal L}_{K}$, $\log{\cal L}_{K} > \log{\cal L}_{\pi}$, $\log{\cal L}_{p} > \log{\cal L}_{\pi}$ and $\log{\cal L}_{e} > \log{\cal L}_{\pi}$.}
    \label{fig:misid}
  \end{center}
\end{figure}
The $K_s$ mass distributions for the four cases we studied are shown in Figure \ref{fig:misid}. The $K_s$ peak is clearly suppressed when the pion is required to be identified as $K$, $p$ or $e$, as expected. 

Overall, the measured identification efficiencies for proton (not presented in this paper), kaon and pion are consistent within a $10\%$ with the Montecarlo expectations, despite the preliminary status of the detector calibration and alignment. Similar discrepancies are also present in the fake rate measurement. We are performing numerous studies to better understand these differences and reduce or eliminate them, as discussed in the next section


\section{Understanding the performances}
On one hand, the early run with the TOP was successful:  the detector took part in more than $90\%$ of the physics runs, with fraction of dead channel of $2.5\%$, and its particle identification capabilities were demonstrated for the first time. On the other hand, the performances are still below the design expectation due to the preliminary status of the calibrations. The TOP calibration consists in a time calibration, whose aim is to even the response of the 8192 MCP-PMT channels, and a geometrical alignment.  The time calibration is performed in four consecutive steps, each one depending on the previous ones \cite{Staric:2017lqu}:

\begin{itemize}
\item Time base calibration. This calibration aims to ensure the linearity of the front-end ASIC sampling array \cite{Kotchetkov:2018qzw}, and is performed injecting electronic pulses in the front-end. After this calibration we measure, using a dedicated laser system \cite{Tamponi:2017dlc}, a single photo-electron time resolution per channel between 100 and 120 ps.

\item Channel T$_0$ calibration. Once the electronics has been properly calibrated, we compensate for the relative delays of the 512 channels within each TOP module. This calibration is performed flashing the MCP-PMTs with a picosecond laser \cite{Tamponi:2017dlc}, and measuring the individual delay of each channel with respect to a reference channel.

\item Geometrical alignment and module T$_0$ calibration. The laser calibration assures that all the delays within each slot are properly compensated, but does not correct for the delays between modules (namely, the relative delays between the reference channels of each module). The synchronization of the modules is expected to be done together with the geometrical alignment by an iterative procedure based on di-muon events from $e^+e^-$ collisions \cite{Staric:2017lqu}. A first attempt to perform a calibration was made, but we found that the $\mu^+ \mu^-$ statistics collected so far is still to small to produce reliable results, and we did not perform any geometrical alignment. However, since the time calibration of the modules with each other is critical for the TOP reconstruction, we developed an alternative algorithm to derive the module T$_0$ using the muons from the 2018 cosmic ray test dataset.

\item Global T$_0$. As mentioned in the introduction, the Cherenkov photon time is measured with respect to the time of the original $e^+e^-$ interaction. A very precise time reference is given by the accelerator radio-frequency (RF) clock, but to use it we need  associate each event with the corresponding bunch crossing. This can be done collecting all the particles detected in the event and, from the time of the hits in the TOP counter, fits the most probable interaction time. This algorithm has a resolution of $\approx 300$ ps (to be compared with the SuperKEKB bunch crossing interval of $2$ ns), corresponding to a bunch crossing identification efficiency greater than $95\%$. For this procedure to be successful, one needs to calibrate the delay between the RF clock and the TOP (or in other words, the the relative phase), to a precision of the order of few tens picoseconds. Any residual phase between the TOP and the RF clock will result in a net extra contribution to the TOP time measurement.  
This calibration is performed separately for each run using di-muon events, but the statistics available still significantly limits its precision to  $\approx 150$ ps on an typical run, $\approx 300$ ps for the shortest runs and  $\approx 30$ ps for the longest ones.
 
\end{itemize}

During and after the data taking, we found several issues in each of these calibration steps. First, residual non-linearities that, despite being rather small, significantly reduced the precision of the laser calibration. Then, the statistics of di-muon collected is not sufficient to perform any of the track-based calibration to the required degree of precision. In some cases, like for the global T$_0$, the statistic resolution represent the largest contribution to the total time resolution of the TOP detector. As mentioned, the geometrical alignment has not been performed, and also the module T$_0$ calibration was done only using cosmic ray events. All these effects, combined with a reduced hit reconstruction efficiency due to the early version the front-end firmware used during phase II, and convoluted with all the effects coming from the preliminary tracking calibration, would fully explain the performance degradation we observed. We'd like to remark that all these problems have known solutions, that will be implemented for the beginning of the Phase III operations, in spring 2019.

\bibliography{rich2018_proceedings}

\end{document}